\documentclass[aps,prx,twocolumn,showpacs,superscriptaddress]{revtex4-2}  
\usepackage{float}
\usepackage{bbm}
\usepackage{epsfig}
\usepackage{epstopdf}
\usepackage{graphicx}
\usepackage{amsmath,amssymb}
\usepackage{amsmath,bm}
\usepackage{physics}
\usepackage{color}
\usepackage{xurl}
\usepackage{lineno,blindtext}
\usepackage{siunitx, booktabs}
\usepackage{diagbox, eqparbox, hhline}
\usepackage{soul}
\usepackage{xcolor}

\usepackage[normalem]{ulem}
\newcolumntype{P}[1]{>{\centering\arraybackslash}p{#1}}
\newcommand{\bbGamma}{{\mathpalette\makebbGamma\relax}}
\newcommand{\makebbGamma}[2]{%
  \raisebox{\depth}{\scalebox{1}[-1]{$\mathsurround=0pt#1\mathbb{L}$}}%
}
\usepackage[colorlinks,citecolor=blue,linkcolor=blue,urlcolor=blue,]{hyperref}

\begin{document}

\title{Solid-state platform for cooperative quantum dynamics \\ driven by correlated emission}

\author{Xin Li}
\email{licqp@bc.edu}
\affiliation{Department of Physics, Boston College, 140 Commonwealth Avenue, Chestnut Hill, Massachusetts 02467, USA}
\author{Jamir Marino}
\affiliation{Institut f$\ddot{\text{u}}$r Physik, Johannes Gutenberg-Universit$\ddot{\text{a}}$t Mainz, D-55099 Mainz, Deutschland}
\author{Darrick E. Chang}
\affiliation{ICFO-Institut de Ciencies Fotoniques, The Barcelona Institute of
Science and Technology, 08860 Castelldefels (Barcelona), Spain}
\affiliation{ICREA-Instituci\'o Catalana de Recerca i Estudis Avan\c{c}ats, 08010 Barcelona, Spain}
\author{Benedetta Flebus}
\affiliation{Department of Physics, Boston College, 140 Commonwealth Avenue, Chestnut Hill, Massachusetts 02467, USA}

\begin{abstract}

While traditionally regarded as an obstacle to quantum coherence, recent breakthroughs in quantum optics have shown that the dissipative interaction of a qubit with its environment can be leveraged to protect quantum states and synthesize many-body entanglement.
Inspired by this progress, here we set the stage for the -- yet uncharted -- exploration of analogous cooperative 
 phenomena in hybrid solid-state platforms. 
We develop a comprehensive formalism for the quantum many-body dynamics of an ensemble of solid-state spin defects interacting  with the magnetic field fluctuations of a common solid-state reservoir. Our framework applies to any solid-state reservoir whose fluctuating spin, pseudospin, or charge degrees of freedom generate magnetic fields. To understand whether correlations induced by dissipative processes can play a relevant role in a realistic experimental setup, we apply our model to a qubit array interacting via the spin fluctuations of a ferromagnetic bath. Our results show that the low-temperature collective relaxation rates of the qubit ensemble can display clear signatures of super- and subradiance, i.e., forms of cooperative dynamics traditionally achieved in atomic ensembles. We find that the solid-state analog of these cooperative phenomena is robust against spatial disorder in the qubit ensemble and thermal fluctuations of the magnetic reservoir, providing a route for their feasibility in near-term experiments.
Our work lays the foundation for a multi-qubit approach to quantum sensing of solid-state systems and the direct generation of many-body entanglement in spin-defect ensembles. Furthermore, we discuss how the tunability of solid-state reservoirs opens up novel pathways for exploring cooperative phenomena in regimes beyond the reach of conventional quantum optics setups.

\end{abstract}

\maketitle

%%%%%%%%%%%%%%%%%%%%%%%%%%%%Main Body%%%%%%%%%%%%%%%%%%%%%%%%%%%%%%%%%%%%%

\section{Introduction}
\label{sec:intro}

While the   coupling between a quantum system and its environment has traditionally been regarded as a hindrance to the development of emerging quantum technologies, recent breakthroughs in the field of light-matter interfaces  have shown that  correlated dissipation acting on an ensemble of qubits   can be harnessed to realize new dynamical states of matter with a broad range of  functionalities~\cite{asenjo2017exponential,chang2014quantum,sipahigil2016integrated,chang2018colloquium,pivovarov2020quantum,reitz2022cooperative,shahmoon2020quantum,bettles2016cooperative,kornovan2016collective,sutherland2016collective,jenkins2012controlled,shahmoon2017cooperative,facchinetti2016storing,marino2022universality,seetharam2022dynamical}.  
In light-matter interfaces,  atomic emitters interact with each other through the emission or absorption of   photons in a common photonic reservoir.
When multiple quantum emitters radiate in the shared bath,  they can correlate giving rise to a form of cooperative relaxation dynamics of the  qubit ensemble, whose emission rate is initially enhanced (\textit{superradiance}) compared to that of an isolated atom~\cite{meiser2009prospects,bohnet2012steady}.  Superradiant states usually decay into  \textit{subradiant} states, i.e.,  many-body states with weak coupling to the environment, whose quantum correlations can be harnessed  for potential applications ranging from quantum-enhanced metrology~\cite{{facchinetti2016storing,ostermann2013protected,manzoni2018optimization}} to quantum-information processing and storage~\cite{verstraete2009quantum,chang2018colloquium,koshino2020protection}. At the same time, the spatial profile of correlated dissipation is relatively constrained in quantum optical setups, and experimental demonstrations to effectively utilize subradiance remain limited to date~\cite{rui2020subradiant,srakaew2023subwavelength}.

While light-matter interfaces have been a foundational platform for exploring quantum cooperative phenomena,  the latter can, in principle, emerge in any ensemble of long-lived quantum systems interacting   with a shared reservoir. Natural candidates are quantum hybrid solid-state systems comprising an ensemble of solid-state spin defects that interact   with the electromagnetic noise emitted by a shared solid-state bath.
%The exploration of quantum cooperative dynamics in alternative platforms has the two-fold potential advantage of pushing forward the capaibilities of these alternative platforms while opening up the pathway for the exploration of these fundamental phenomena in regimes that are beyond  the reach of quantum optics setups (new phenomena can be harnesssed). 
\begin{figure*}[htbp]
    \centering
\includegraphics[width=1\linewidth]{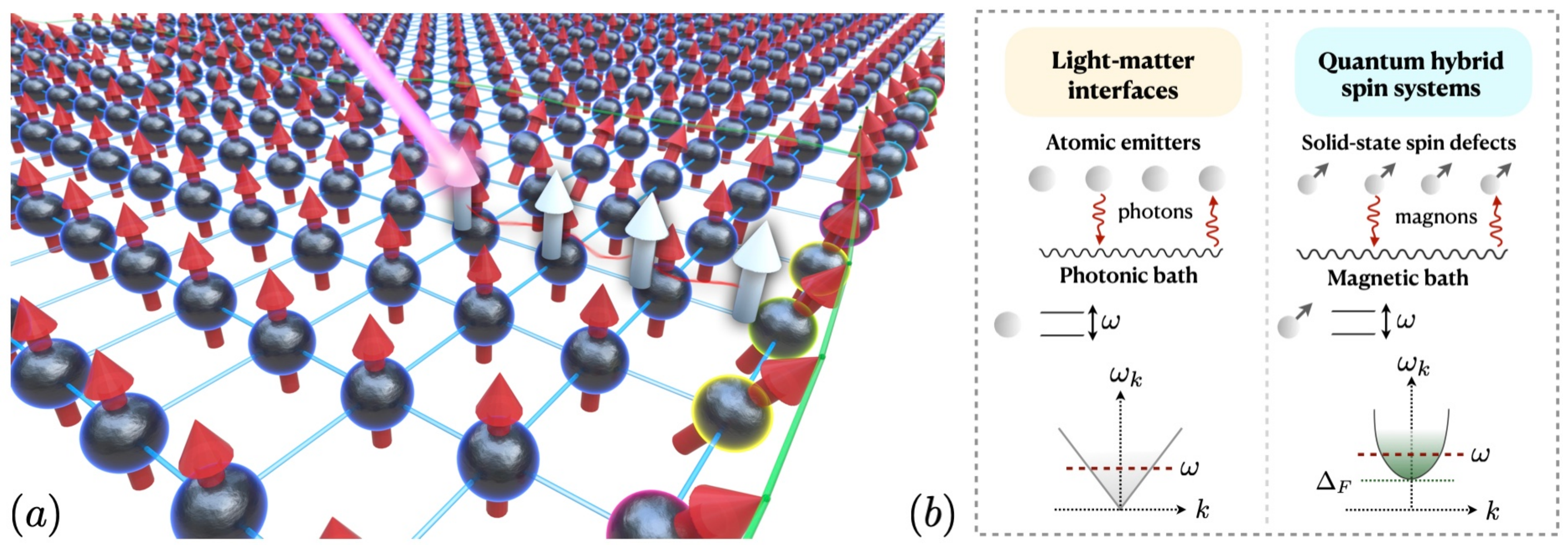}
    \caption{(a) Qubit-qubit correlations mediated by the spin waves of a magnetic system. The blue square lattice represents the magnet, with the red arrows indicating the ground state spin polarization and the green lines highlighting the spin-wave fluctuations. The spin qubits (grey arrows) are arranged in a $1d$ array  in a parallel plane above the magnetic lattice. (b) Schematic illustration of the minimal ingredients to realizing cooperative quantum dynamics, i.e.,  an ensemble of long-lived quantum systems interacting with a common  bath. At light-matter interfaces, atomic emitters interact with each other via a shared photonic bath with dispersion $\omega_{k}$. Since the resonance frequency $\omega$ of the quantum emitters is within the gapless photonic continuum, the quantum emitters can emit (absorb) \textit{real photons} into (from) the bath. An analogous regime can be realized in quantum hybrid spin systems when the frequency $\omega$ of the solid-state spin defects is larger than the gap $\Delta_{F}$ of the spin-wave dispersion $\omega_{k}$ of the common magnetic reservoir.}
    \label{Fig1}
\end{figure*}
The dissipative dynamics of    single spin defects 
are routinely leveraged in     state-of-the-art quantum sensing approaches, which probe the solid-state bath properties via their effect on the sensor spin's relaxation time~\cite{casola2018probing}. However, to date, solid-state spin defects have  been essentially operated as \textit{single-qubit sensors}. Sensing schemes that explore noise cross-correlations between a couple of nitrogen-vacancy (NV) centers have only been implemented at temperatures high enough to suppress any potential bath-driven  quantum correlations among the sensors~\cite{rovny2022nanoscale}.    
%Even just at a theoretical level, a framework for multi-qubit spectroscopy of spatio-temporally correlated quantum noise generated by solid-state reservoirs is lacking. 
To this end, a key challenge remains in engineering robust long-distance quantum correlations between nitrogen-vacancy (NV) centers~\cite{dolde2013room,bermudez2011electron,gaebel2006room,sipahigil2012quantum,bernien2013heralded,pfaff2014unconditional,hensen2015loophole,reiserer2016robust,degen2017quantum}, which could enable the scaling of spin-defect-based platforms for a wide array of applications, including not only quantum sensing but also quantum computing and communication networks.

Building on previous works that explore magnetic systems as intermediary of coupling among distant NV centers~\cite{trifunovic2013long,andrich2017long,flebus2019entangling,muhlherr2019magnetic,zou2020tuning,gonzalez2022towards}, a recent proposal by Fukami \emph{et al.}~\cite{fukami2021opportunities} suggests leveraging the interactions between NV centers and a YIG waveguide to achieve two-qubit NV interactions. This approach is anticipated to enable high-fidelity entanglement over micrometer-scale distances by utilizing both off- and on-resonant magnon exchange at temperatures below 150 mK. Our work aims to scale these platforms to the \emph{many-body regime} by demonstrating how they can support cooperative relaxation dynamics, a phenomenon traditionally studied in quantum optics~\cite{gross1982superradiance,reitz2022cooperative}.
The key resource in our approach is the spatio-temporally correlated noise emitted by a common reservoir, whose impact on the quantum dynamics of two qubits has been explored across various solid-state platforms~\cite{von2020two,yoneda2023noise,rojas2023spatial}.  Our overarching vision is to transfer the field of dissipative state engineering from atomic, molecular, and optical (AMO) physics to quantum hybrid solid-state systems~\cite{van2024dissipation,yuan2022quantum}. There is a growing body of theoretical research suggesting that dark states of Lindblad dynamics could be used to synthesize few- or many-body entangled states, effectively transforming decoherence from a hindrance into a resource~\cite{diehl2008quantum,marino2016driven,ma2019dissipatively,verstraete2009quantum}. However, while concrete implementations in AMO systems remain scarce, a structured dissipation framework for many-body systems in solid-state platforms has yet to be fully developed. 
This framework would also naturally lay the foundation for developing quantum sensing techniques that harness quantum correlations among solid-state spin defects -- an area that remains largely unexplored.

In this work, we propose a concrete experimental route to realize the dynamical crossover from super- to subradiance using correlated dissipation acting on an ensemble of NV centers.
The theoretical framework we develop is applicable to a wide range of solid-state spin defects interacting with any  solid-state reservoir  whose fluctuating degrees of freedom can generate magnetic fields. Candidates include  spin fluctuations of magnetically ordered systems~\cite{van2015nanometre,du2017control,wang2022noninvasive,yan2022quantum}, electric current fluctuations in metallic and superconducting materials~\cite{chatterjee2022single,agarwal2017magnetic}, and valley-polarized currents in Weyl semimetals~\cite{zhang2022flavors}. By demonstrating the possibility to realize the most iconic forms of cooperative relaxation phenomena  in NV ensembles, we open the door to further, more elaborate, dissipative preparations of correlated many-body states in the field of solid-state quantum information science. 
A boon  of this approach lies in the ability to enhance the effective qubit-bath interaction by increasing the size of the qubit ensemble (following   quantum optics jargon, this is   called \emph{cooperative} enhancement~\cite{reitz2022cooperative}).
Since the range of correlated dissipation falls with long tails, qubits can radiate in synchronous blocks, offering thereby protection against experimental perturbations (such as inhomogeneous broadening or positional disorder), as the qubit ensemble scales to large sizes. 

 %Therefore, compared with the binary operations implemented in traditional two-qubit gate architectures, simultaneously acting, long-range interactions can significantly streamline quantum circuit designs
%and accelerate their execution,  thereby reducing the effects of decoherence~\cite{katz2022n}. 

To understand whether dissipative correlations can play a relevant role in a realistic experimental setup, we apply our general framework to the quantum hybrid spin system shown in Fig.~\ref{Fig1}(a): an  NV-center array with lattice constant $a$ interacting coupled to a  ferromagnetic bath, i.e., a Yttrium Iron Garnet (YIG) film.  
As illustrated in Fig.~\ref{Fig1}(b), this setup allows us to draw a clear analogy with the mechanisms driving quantum cooperative phenomena in quantum optics. At light-matter interfaces,  atomic emitters interact  with a common photonic bath with  dispersion $\omega_{k}$. The gapless nature of the photonic reservoir allows for   the emission (absorption) of \textit{real photons} into (from) the bath, at any qubit resonance frequency $\omega$.  An alike exchange of \textit{real magnons} between each NV center and the ferromagnetic bath is enabled by tuning the NV-center resonance $\omega$ frequency  \textit{above  the gap} $\Delta_{F}$  of the YIG spin-wave dispersion. It is worth noting that this mechanism has been already leveraged to extract statistical and transport properties of YIG spin waves from the relaxation rates of an ensemble of \textit{uncorrelated} NV centers, i.e., operated at  temperatures high enough to  suppress any potential qubit-qubit quantum correlations~\cite{du2017control,wang2022noninvasive}. 

Our results show that the relaxation dynamics of an NV-center ensemble coupled to a YIG film can display clear signatures of cooperative quantum behavior, i.e., superradiant and subradiant dynamics, when the system is cooled down to the quantum regime. We perform our calculations within a parameter regime relevant to near-term NV-center-based experiments and find that the qubit-qubit correlations induced by the ferromagnetic bath are remarkably robust against spatial disorder and thermal fluctuations even in a small ($N=9$) NV-center ensemble. A comparison with recent experimental findings suggests that many-body quantum correlations in denser  ensembles may be resilient against significant inhomogeneous broadening of the qubit resonances~\cite{angerer2018superradiant}, benefiting of the   cooperative enhancement mentioned above~\cite{putz2014protecting,norcia2018cavity,reitz2022cooperative,masson2020many}.

Crucial differences in the physical properties of magnons and photons open up a new realm of functionalities beyond the reach of quantum optics setups.
For instance, while the long wavelengths of photons make them suitable for large systems, magnons can offer a compact platform for on-chip nano-devices. Photons are non-interacting, gapless, linearly dispersing bosonic excitations. Magnons, instead, are intrinsically interacting quasi-particles that display tunable band gaps and diverse dispersions owing to the rich plethora of magnetic materials. 
Furthermore, while the continuum of electromagnetic modes responsible for phenomena like super- and subradiance constitutes an inherent entity, magnetic baths, as well as solid-state systems in general, can be steered by currents, thermal biases, and a range of other non-equilibrium driving knobs \cite{cornelissen2016magnon,uchida2008observation,barman20212021}.
Thus, our framework naturally paves the way for studying cooperative phenomena and structured dissipation   in the presence of far-from-equilibrium baths that lack counterparts in light-matter interfaces. We further elaborate on these directions  in the concluding section of the manuscript.

 This work is organized as follows. In Sec.~\ref{sec:model}, we develop a general theoretical framework for the quantum many-body dynamics of an array of solid-state spin defects interacting with the magnetic field fluctuations of a common solid-state reservoir. In Sec.~\ref{sec:bath}, we introduce a model for the ferromagnetic bath. In Sec.~\ref{sec:ME}, we apply the formalism developed in Sec.~\ref{sec:model} to the ferromagnetic bath  and investigate the effective qubit-qubit interactions in experimentally relevant regimes. In Sec.~\ref{sec:single}, we explore the dynamics of the quantum hybrid spin system in the lowest Dicke, i.e., single-excitation,  manifold. In Sec.~\ref{sec:multi}, we investigate the quantum many-body dynamics of the ensemble and its robustness to experimental perturbations.
We provide a summary and outlook in Sec.~\ref{sec:concl}.

\section{Model}
\label{sec:model}

In this section, we explore the quantum many-body dynamics of an array of solid-state spin defects interacting with a common solid-state reservoir.  The starting point of our work is a one-dimensional (1$d$) array of solid-state spin defects, modeled as two-level systems, which couple to the magnetic field $\textbf{B}$ generated by a nearby reservoir. The Hamiltonian of the (isolated) qubit array reads as
\begin{align}
\mathcal{H}_s = -\frac{1}{2}\sum_\alpha \omega_{\alpha}\boldsymbol{\sigma}_\alpha\cdot\hat{\mathbf{n}}_\alpha\,,
\end{align}
where 
$\hat{\mathbf{n}}_\alpha = (\sin\theta_\alpha\cos\phi_\alpha, \sin\theta_\alpha\sin\phi_\alpha, \cos\theta_\alpha)$ labels the equilibrium orientation of the quantum spin $\boldsymbol{\sigma}_\alpha$ with resonance frequency  $\omega_{\alpha}$  residing at the site $\mathbf{r}_{\alpha}$.
The  Zeeman interaction between the qubit array and a magnetic field can be generally written as

\begin{align}
\mathcal{H}_{int}=-\tilde{\gamma}\sum_\alpha(B_\alpha^+\sigma^-_\alpha+B_\alpha^-\sigma^+_\alpha+B_\alpha^z\sigma^z_\alpha ),\label{bathfield}
\end{align} 
 where $\tilde{\gamma}$ is the gyromagnetic ratio of the 
 solid-state spin defects. The components $B_\alpha^{\pm} = 
 B_\alpha^x \pm iB_\alpha^y$ and $B_\alpha^z$ correspond to 
 the local magnetic field experienced by the $\alpha$th spin, while $\sigma^\pm_\alpha = (\sigma^x_\alpha \pm 
 i\sigma^y_\alpha)/2$ and $\sigma^z_\alpha$ are the Pauli 
 matrices expressed in the local coordinates of the $
 \alpha$th qubit.

%
%In addition to the challenges posed by the numerous degrees of freedom in magnetic materials, symmetries also play a crucial role in determining the physical properties of the system. These symmetries can offer valuable insights and aid in simplifying and modifying the form of the master equation. We will  show that, by exploiting the relevant symmetries present in the system, e.g., $U(1)$ symmetry in the spin space and rotational symmetry in the real space, we can uncover certain patterns and relationships that enable further simplifications and analytical solutions.
%
%We provide a demonstration that the coefficients present in the master equation can be correlated with the spin densities of the magnetic bath through the utilization of magnetostatic Green's functions. By employing the linear response theory and dissipation-fluctuation theorem, we clarify that, under specific conditions, all the coupling coefficients can be expressed in terms of the susceptibilities derived from the LLG equation and the diffusion equation, thus enabling the magnetic bath to determine the properties of the MSQs.
The  interactions between a solid-state spin defect and the field fluctuations of a nearby solid-state reservoir are much weaker than the characteristic energy scales  of the individual systems.  Thus, in the Schr\"{o}dinger picture, the evolution of the 
density matrix $\rho$ of the array can be generally  
described by the following Markovian master equation~\cite{zou2022bell,breuer2002theory,weiss2012quantum,gardiner2004quantum,rivas2012open,davies1974markovian,gorini1978properties}: 
\begin{align}
    \frac{d\rho}{dt}= -i \left[ \mathcal{H}_s+\mathcal{H},\;\rho \right] + \mathcal{L}\left[ \rho\right],\label{densitymatrix}
\end{align}
where the Hamiltonian $\mathcal{H}$ reads as 
\begin{align}
\mathcal{H}=  \sum_{\alpha,\beta}\sum_{(\mu,\tilde{\mu})}\sum_{(\nu,\tilde{\nu})}Y^{\mu\nu}_{\alpha\beta} \sigma_\alpha^{\tilde{\mu}}\sigma_\beta^{\tilde{\nu}},\label{master:hamil}
\end{align}
and
\begin{align}
\mathcal{L}\left[ \rho\right]= \sum_{\alpha,\beta}\sum_{(\mu,\tilde{\mu})}\sum_{(\nu,\tilde{\nu})}X_{\alpha\beta}^{\mu\nu}\left( \sigma_\beta^{\tilde{\nu}}\rho\sigma_\alpha^{\tilde{\mu}}-\frac{1}{2}\{ \sigma_\alpha^{\tilde{\mu}}\sigma_\beta^{\tilde{\nu}},\rho\}\right),\label{master:linb}
\end{align}
is the Lindbladian.
Here, the sums over $(\mu, \tilde{\mu})$ and $(\nu, \tilde{\nu})$ are taken over the combinations $(\mu, \tilde{\mu}) = (\pm, \mp), (z, z)$, where $\pm = x \pm i y$. The explicit forms of $Y_{\alpha\beta}$ and $X_{\alpha\beta}$ are given by
\begin{align}
Y^{\mu\nu}_{\alpha\beta}&=\frac{\tilde{\gamma}^2}{2}\int_0^\infty d\tau \; G^>_{B_{\alpha}^\mu B_{\beta}^\nu}(\tau) e^{-i\omega_\beta^\nu\tau}\nonumber\\
&-\frac{\tilde{\gamma}^2}{2}\int_0^\infty d\tau \; G^<_{B_{\beta}^\nu B_{\alpha}^\mu }(\tau) e^{-i\omega_\alpha^{\mu}\tau},\label{coherentcop}\\
X^{\mu\nu}_{\alpha\beta}&=i\tilde{\gamma}^2\int_0^\infty d\tau \; G^>_{B_{\alpha}^\mu B_{\beta}^\nu}(\tau) e^{-i\omega_\beta^\nu\tau}\nonumber\\
&+i\tilde{\gamma}^2\int_0^\infty d\tau \; G^<_{B_{\beta}^\nu B_{\alpha}^\mu }(\tau) e^{-i\omega_\alpha^{\mu}\tau},\label{decoherentcop}
\end{align}
where the relevant frequencies are defined as $\omega_{\alpha(\beta)}^\pm=\pm\omega_{\alpha(\beta)}$ and $\omega_{\alpha(\beta)}^z=0$, and 
the greater and lesser Green’s functions follow the conventional definition~\cite{haug2008quantum} for two general operators $\mathcal{O}_1$ and $\mathcal{O}_2$:
\begin{align}
G^>_{\mathcal{O}_1,\mathcal{O}_2}(\tau)\equiv-i\langle \mathcal{O}_1(\tau)\mathcal{O}_2\rangle,\\ G^<_{\mathcal{O}_1,\mathcal{O}_2}(\tau)\equiv-i\langle \mathcal{O}_2\mathcal{O}_1(\tau)\rangle.
\end{align}

The  term $\propto Y_{\alpha \beta}$~\eqref{coherentcop}  describes \textit{coherent} qubit-qubit interactions that conserve the number of excitations, while the term $\propto X_{\alpha \beta}$~\eqref{decoherentcop}  represents the generator of 
 the \textit{incoherent dynamics}. For an isolated quantum spin, i.e., $\alpha=\beta$, the  terms 
on the right-hand side of Eq.~(\ref{master:linb}) reduce to the \textit{single-qubit} relaxation and 
dephasing rates routinely measured in quantum 
sensing experiments~\cite{casola2018probing}.  Incoherent processes  also have a \textit{collective} character encoded in the off-diagonal elements of  $X$ (i.e., for $\alpha \neq \beta$), i.e., the noise cross-correlations invisible to single-qubit sensors  that are known to generate  cooperative dynamics and entanglement in light-matter interfaces~\cite{diehl2008quantum,asenjo2017exponential,verstraete2009quantum}.

 The formalism we have introduced here is rather general as it is applicable to any bath whose fluctuating spin, pseudospin or charge degrees of freedom generate magnetic fields. In the following,  we focus on a qubit array interacting via long-range dipole-dipole interactions with a magnetically ordered bath, as shown in Fig.~\ref{Fig1}(a). Fluctuations of the spin density $\mathbf{s}_\mathbf{r}$ of the magnetic bath generate a stray field 
 \begin{align}
\mathbf{B}_\alpha =  \int d^n \mathbf{r} \; \mathcal{D}_{\mathbf{r}_\alpha, \mathbf{r}} \; \mathbf{s}_\mathbf{r}, \label{3dmagnetic}
\end{align}
where $\mathcal{D}_{\mathbf{r}_\alpha,\mathbf{r}}$ is the tensorial Green’s function of the magnetic dipolar field, which can be obtained as solutions of the Maxwell's equations   in the magnetostatic limit~\cite{guslienko2011magnetostatic}. Using Eq.~\eqref{3dmagnetic}, the greater (lesser) Green’s functions in Eqs.~\eqref{coherentcop} and~\eqref{decoherentcop} can be written as

\begin{eqnarray}
G^{>,(<)}_{B^{\mu}_{\alpha}B^{\nu}_{\beta}}(\tau)=\iint  d^n\mathbf{r}  \;  d^n\mathbf{r}' \; \mathcal{D}^{\mu\mu'}_{\mathbf{r}_\alpha,\mathbf{r}}\mathcal{D}^{\nu\nu'}_{\mathbf{r}_\beta,\mathbf{r}'}G^{>,(<)}_{s^{\mu'}_{\mathbf{r}}s^{\nu'}_{\mathbf{r}'}}(\tau),\label{gengl}
\end{eqnarray}
with $\mu,\nu=\pm,z$, and where the repeated indices represent the Einstein summation. In the following, we  discuss how symmetries that are relevant to common experimental realizations of the proposed setup allow to simplify the form of the master equation~\eqref{densitymatrix}. 

The magnetostatic Green's function $\mathcal{D}^{\mu\nu}_{\mathbf{r}_\alpha,\mathbf{r}}$ (\ref{3dmagnetic}) represents the response of the magnetic bath at position $\mathbf{r}_\alpha$ to an external perturbation at position $\mathbf{r}$.
For a translationally-symmetric bath, the magnetostatic Green's function only depends on the relative position vector $\mathbf{r}_\alpha-\mathbf{r}$,  i.e.,  $\mathcal{D}^{\mu\nu}_{\mathbf{r}_\alpha,\mathbf{r}}=\mathcal{D}^{\mu\nu}_{{\mathbf{r}_\alpha-\mathbf{r}}}$. Analogously, we can set $G^{>,(<)}_{s^{\mu'}_{\mathbf{r}}s^{\nu'}_{\mathbf{r}'}}(\tau) = G^{>,(<)}_{s^{\mu'}s^{\nu'}}(\tau,\mathbf{r}-\mathbf{r}')$.  
We can then substitute the Fourier transform of the magnetostatic Green's function, i.e.,\begin{equation}
\mathcal{D}^{\mu\nu}_{\mathbf{r}_\alpha-\mathbf{r}}= \int \frac{d^n\mathbf{k}}{(2\pi)^n} e^{i\mathbf{k}\cdot(\mathbf{r}_\alpha-\mathbf{r})}\mathbb{D}^{\mu\nu}_\mathbf{k} ,
\end{equation}
into Eq.~(\ref{gengl}), which leads to
\begin{align}
G^{>,(<)}_{B^{\mu}_{\alpha}B^{\nu}_{\beta}}(\tau)=\int  \frac{d^n\mathbf{k}}{(2\pi)^n}e^{i\mathbf{k}\cdot \mathbf{r}_{\alpha \beta}}\mathbb{G}^{\mu\nu}(\tau,\mathbf{k}),\label{suscepmomspace}
\end{align}
where $\mathbf{r}_{\alpha \beta}=\mathbf{r}_\alpha-\mathbf{r}_\beta$ is the displacement between  the quantum spins $\boldsymbol{\sigma}_{\alpha}$ and $\boldsymbol{\sigma}_{\beta}$, and 

\begin{align}
    \mathbb{G}^{\mu\nu}(\tau,\mathbf{k})=\mathbb{D}^{\mu\mu'}_{\mathbf{k}}\mathbb{D}^{\nu\nu'}_{-\mathbf{k}}\mathbb{G}^{>,(<)}_{s^{\mu'}s^{\nu'}}(\tau,\mathbf{k}),
\end{align}
where $\mathbf{k}$ is the wavevector in Fourier space and $\mathbb{D}^{\mu\nu}_{\mathbf{k}}$ and $\mathbb{G}^{>,(<)}_{s^{\mu'}s^{\nu'}}(\mathbf{k})$ represent the Fourier-transformed Green's functions.  Moreover, isotropicity in the real space allows to further simplify $G^{>,(<)}_{s^{\mu'}s^{\nu'}}(\tau,\mathbf{r}-\mathbf{r}')=G^{>,(<)}_{s^{\mu'}s^{\nu'}}(\tau,|\mathbf{r}-\mathbf{r}'|)$.

Another symmetry relevant to several magnetic reservoirs is $U(1)$ symmetry, i.e., the symmetry under spin rotations around the direction of magnetic field (here taken to be along the $\hat{\mathbf{z}}$ axis), which implies the conservation of the spin projection on the direction of the magnetic field, and, thus, of the magnon number.
While the magnon number can not be an exactly conserved quantity due to the ubiquitous dissipative interactions between magnons and the crystalline lattice, it is generally taken to be \textit{approximately} conserved in  collinear magnetic systems that display low damping and strong number-conserving interactions \cite{flebus2021magnonics}. This assumption allows us to simplify our model further as the transverse spin response is completely decoupled from the longitudinal spin dynamics, i.e., $\langle \hat{s}^{\pm}(t,\mathbf{r})\hat{s}^{z}(0,\mathbf{r}') \rangle=0$ and $\langle \hat{s}^{\pm}(t,\mathbf{r})\hat{s}^{\pm}(0,\mathbf{r}')\rangle=0$. 
When the (equilibrium) spin density of the magnetic bath and the quantum spins are collinear, i.e., $\theta_{\alpha}=0$ for $\alpha=1,\cdots,N$, we also find that only two-point correlations of the form  $\langle B^-B^+\rangle$, $\langle B^+B^-\rangle$, and $\langle B^zB^z\rangle$ are non-vanishing.

Under these assumptions, the  Hamiltonian (\ref{master:hamil}) takes the form of an  XXZ model with long-range couplings, i.e.,
 \begin{eqnarray}
\mathcal{H}=\sum_{
\alpha\neq\beta}J_{\alpha\beta}\sigma_\alpha^+\sigma_\beta^-+\sum_{
\alpha\neq \beta}J^{z}_{\alpha\beta}\sigma_\alpha^z\sigma_\beta^z,\label{uoneham}
\end{eqnarray}
with 
\begin{eqnarray}
J_{\alpha\beta}
&=&\frac{\tilde{\gamma}^2}{2}\left[G^R_{B^-_{\alpha}B^+_{\beta}}(-\omega_{\beta})+G^R_{B^+_{\beta}B^-_{\alpha}}(\omega_{\alpha})\right],\label{perpJ}\\
J^{z}_{\alpha\beta}
&=&\lim_{\omega\rightarrow0^+}\frac{\tilde{\gamma}^2}{4}\left[G^R_{B^z_{\alpha}B^z_{\beta}}(\omega)+G^R_{B^z_{\beta}B^z_{\alpha}}(\omega)\right],\label{paraJ}
\end{eqnarray}
where $0^+$ in the limit $\lim_{\omega\rightarrow0^+}$ represents an infinitesimal but finite frequency and we have neglected the overall  energy shift proportional to $\sum_\alpha(Y_{\alpha\alpha}^{-+}+Y_{\alpha\alpha}^{+-}+2Y_{\alpha\alpha}^{zz})\sigma_\alpha^0/2$ and the Lamb shift proportional to $\sum_\alpha(Y_{\alpha\alpha}^{-+}-Y_{\alpha\alpha}^{+-})\sigma_\alpha^z/2$. Here the retarded Green's function for two general operators $\mathcal{O}_1$ and $\mathcal{O}_2$  is defined as

\begin{eqnarray}
    G^R_{\mathcal{O}_1\mathcal{O}_2}(\tau)=-i\Theta(\tau)\langle [\mathcal{O}_1(\tau),\mathcal{O}_2]\rangle, 
    \label{green}
\end{eqnarray}
where $\Theta(\tau)$ is the Heaviside step function.
The Lindbladian operator  ($\ref{master:linb}$) simplifies to
 \begin{eqnarray}
 \mathcal{L}[\rho]&&=\sum_{\alpha\beta}\Gamma_{\alpha\beta}\left(\sigma_\beta^+\rho\sigma_\alpha^--\frac{1}{2}\{\sigma_\alpha^-\sigma_\beta^+,\rho\}\right)\nonumber\\
 &&+\sum_{\alpha\beta}\tilde{\Gamma}_{\alpha\beta}\left(\sigma_\beta^-\rho\sigma_\alpha^+-\frac{1}{2}\{\sigma_\alpha^+\sigma_\beta^-,\rho\}\right)\nonumber\\
 &&+\sum_{\alpha\beta}\Gamma^z_{\alpha\beta}\Big(\sigma_\beta^z\rho\sigma_\alpha^z-\frac{1}{2}\{\sigma_\alpha^z\sigma_\beta^z,\rho\}\Big),\label{uonelinb}
 \end{eqnarray}
 where, for notational convenience, we have redefined $\Gamma_{\alpha\beta}=X_{\alpha\beta}^{+-}$,  
$\tilde{\Gamma}_{\alpha\beta}=X_{\alpha\beta}^{-+}$ and $\Gamma^z_{\alpha\beta}=X_{\alpha\beta}^{zz}$.

Equation~\eqref{uonelinb}  encodes structured dissipation in the form of (i) correlated emission ($\Gamma_{\alpha\beta}$), which describes how the emission of a magnon from one qubit into the environment influences the likelihood of magnon emission from another qubit in the ensemble; (ii) correlated pumping  ($\tilde{\Gamma}_{\alpha\beta}$), i.e., the incoherent absorption of magnons from the material, and  (iii) correlated dephasing ($\Gamma^z_{\alpha\beta}$), which arises from noisy Markovian magnetic fields longitudinal to the qubits quantization axis that display a spatially correlated pattern along the whole qubit ensemble. The first term is responsible for super- and subradiance and is broadly discussed in  AMO literature, as outlined in Sec.~\ref{sec:intro}. 
In the following, we will discuss how to attenuate or suppress the second and third dissipative channels in order to achieve a clean fingerprint of collective emission in NV-center ensembles. However, it is essential to note that these terms are easily restorable and, importantly, tunable in solid-state baths driven out of equilibrium. This is in stark contrast with light-matter interfaces, where these forms of correlated pumping and dephasing would be difficult to realize.
We will further comment on how these crucial differences can enable the exploration of novel forms of dissipative many-body state preparation exclusive of our solid-state setup in Sec.~\ref{sec:concl}.

For an ensemble of qubits with identical transition frequencies, denoted as $\omega_{qi}=\omega_\alpha$ for $\alpha=1,\cdots,N$, the coefficients in the Lindbladian (\ref{uonelinb}) can be further simplified to
\begin{eqnarray}
\Gamma_{\alpha\beta}&=&i\tilde{\gamma}^2G^{>}_{B_\alpha^+B_\beta^-}(\omega_{qi}),\label{decoem}\\
\tilde{\Gamma}_{\alpha\beta}&=&i\tilde{\gamma}^2G^{<}_{B^+_\beta B^-_\alpha}(\omega_{qi}),\label{decoab}\\
\Gamma^z_{\alpha\beta}&=&\lim_{\omega\rightarrow0^+}i\tilde{\gamma}^2G^{>}_{B_\alpha^zB_\beta^z}(\omega).\label{decozz}
\end{eqnarray}
For an individual qubit, i.e., $\alpha=\beta$, the  coefficients $ \Gamma_{\alpha\alpha}\equiv \Gamma_0$ and  $ \tilde{\Gamma}_{\alpha\alpha}\equiv \tilde{\Gamma}_0$  govern the local relaxation rate,  while  $\Gamma^z_{\alpha\alpha}\equiv \Gamma^z_0$ reduces to the local dephasing rate.
In thermal equilibrium, the fluctuation-dissipation theorem relates Eqs.~(\ref{decoem}) and (\ref{decoab}) as $\tilde{\Gamma}_{\alpha\beta}=e^{-\beta h\omega_{\text{qi}}}\Gamma_{\alpha\beta}$, where $\beta=1/(k_BT)$, $k_B$ is the Boltzmann constant, $T$ is the temperature, and $h$ is the Planck's constant~\cite{kubo1966fluctuation}. 
The term $\propto \tilde{\Gamma}_{\alpha\beta}$ describes the spin-qubit absorption of spin angular momentum from the thermal reservoir: at zero temperature, this term vanishes since there are no thermally populated magnons in the film.

Invoking the fluctuation-dissipation theorem allows to rewrite Eqs.~\eqref{decoem} and~\eqref{decozz} as 
\begin{align}
\Gamma_{\alpha\beta}&=\int  \frac{d^n\mathbf{k}}{(2\pi)^n} e^{i\mathbf{k}\cdot \mathbf{r}_{\alpha \beta}}\bbGamma(\omega_{qi},\mathbf{k}), \label{dissirealspace}\\
\Gamma^z_{\alpha\beta}&=\lim_{\omega\rightarrow0^+}\int  \frac{d^n\mathbf{k}}{(2\pi)^n}e^{i\mathbf{k}\cdot \mathbf{r}_{\alpha \beta}}\bbGamma^z(\omega,\mathbf{k}),\label{dephaserealspace}
\end{align}
where $\bbGamma(\omega_{qi},\mathbf{k})$ and $\bbGamma^z(\omega,\mathbf{k})$ represent $\Gamma_{\alpha\beta}$ and $\Gamma^z_{\alpha\beta}$ in momentum space, respectively, and are given by
\begin{align}
&\bbGamma(\omega_{qi},\mathbf{k})=2\tilde{\gamma}^2[n_B(\omega_{qi})+1]\mathbb{D}^{+\mu}_{\mathbf{k}}\mathbb{D}^{-\tilde{\mu}}_{-\mathbf{k}}\text{Im}[\chi^{\mu\tilde{\mu}}(\omega_{qi},\mathbf{k})],\label{disscoupling}\\
&\bbGamma^z(\omega,\mathbf{k})\,=2\tilde{\gamma}^2[n_B(\omega)+1] \mathbb{D}^{z\mu}_{\mathbf{k}}\mathbb{D}^{z\tilde{\mu}}_{-\mathbf{k}}  \text{Im}[\chi^{\mu\tilde{\mu}}(\omega,\mathbf{k})].
\end{align}
Here, the repeated indices 
$\mu$ and $\tilde{\mu}$
imply Einstein summation, with $(\mu, \tilde{\mu}) = (\pm, \mp), (z, z)$.  Im[$  \chi^{\mu\nu}(\omega_{qi},\mathbf{k})$] denotes the imaginary part of the dynamical spin susceptibility tensor $\chi$, which is defined as
\begin{eqnarray}
    \chi^{\mu\nu}(\omega_{qi},\mathbf{k})=-\,\mathbb{G}^R_{s^\mu s^\nu}(\omega_{qi},\mathbf{k}),
    \label{267}
\end{eqnarray}
where $\mathbb{G}^R_{s^\mu s^\nu}(\omega_{qi},\mathbf{k})$ is the retarded Green's function in momentum space. Similarly, we can rewrite the coherent couplings~(\ref{perpJ}) and (\ref{paraJ}) as
\begin{align}
J_{\alpha\beta}
&=\int  \frac{d^n\mathbf{k}}{(2\pi)^n}e^{i\mathbf{k}\cdot \mathbf{r}_{\beta\alpha }}\mathbb{J}(\omega_{qi},\mathbf{k}),\label{coherealspace}\\
J^{z}_{\alpha\beta}
&=\lim_{\omega\rightarrow0^+}\int  \frac{d^n\mathbf{k}}{(2\pi)^n}e^{i\mathbf{k}\cdot \mathbf{r}_{\beta\alpha }}\mathbb{J}^z(\omega,\mathbf{k}),\label{dephaserealspace}
\end{align}
where $\mathbb{J}(\omega_{qi},\mathbf{k})$ and $\mathbb{J}^z(\omega,\mathbf{k})$ represent $J_{\alpha\beta}$ and $J^z_{\alpha\beta}$ in momentum space, respectively,
\begin{align}
  \mathbb{J}(\omega_{qi},\mathbf{k})\,=&-\tilde{\gamma}^2\,\mathbb{D}^{+\mu}_{\mathbf{k}}\mathbb{D}^{-\tilde{\mu}}_{-\mathbf{k}}\text{Re}[\chi^{\mu\tilde{\mu}}(\omega_{qi},\mathbf{k})],\label{Jcoupling}\\
  \mathbb{J}^z(\omega,\mathbf{k})\,=&-\tilde{\gamma}^2\,\mathbb{D}^{z\mu}_{\mathbf{k}}\mathbb{D}^{z\tilde{\mu}}_{-\mathbf{k}}\text{Re}[\chi^{\mu\tilde{\mu}}(\omega,\mathbf{k})]/2,
\end{align}
where $\text{Re}[\chi^{\mu\nu}(\omega_{qi},\mathbf{k})]$ represents the real part of the dynamical spin susceptibility tensor $\chi$.

\section{Ferromagnetic bath}
\label{sec:bath}

We will now apply the framework developed in the previous section to explore the properties of a qubit array interacting via a simple common magnetic reservoir, i.e., a $U(1)$-symmetric homogeneous ferromagnetic film. The key quantity controlling the effective qubit-qubit coupling is the dynamical spin susceptibility $\chi$ of the magnetic bath, which describes its
linear response  to an externally applied time-dependent perturbative field  $\mathbf{h}$. The susceptibility  $\chi$ is a  tensor that, in general, couples  all components of the spin response to each other. However, for a $U(1)$-symmetric system,  the full response decouples into a longitudinal and transverse part.   The transverse response can be compactly described using the complex variables  $s^{\pm}=( s^{x}\pm i s^{y})/2$ and  $h^{\pm}=h^x \pm ih^y$, yielding for the $-$ component
\begin{eqnarray}
   s^{-}(\omega, k)=\chi^{-+}(\omega, k)\gamma h^{-}(\omega, k),
   \label{sminus}
\end{eqnarray}
where $\gamma$ is the gyromagnetic ratio and $\chi^{+-}(\omega,k)=\left[ \chi^{-+}(-\omega,-k)\right]^*$. Similarly, the longitudinal spin susceptibility  $\chi^{zz}(\omega,k)$ can be introduced as
\begin{eqnarray}
   s^{z}(\omega, k)=\chi^{zz}(\omega, k)\gamma h^{z}(\omega, k).
     \label{sz}
\end{eqnarray}

Using Eq.~(\ref{267}) and introducing the Holstein-Primakoff transformation~\cite{holstein1940field}, i.e., $s^{z}=s - a^{\dagger} a$ and $s^{-} \simeq \sqrt{2s}  a^{\dagger}$, the transverse~\eqref{sminus} and longitudinal~\eqref{sz} spin susceptibilities can be straightforwardly related to, respectively,  two-point and four-point magnon correlation functions. The transverse spin susceptibility~\eqref{sminus} is proportional to the magnon density of states. Thus, its contribution to the dissipative qubit-qubit interactions~\eqref{disscoupling}  is associated with one-magnon processes, i.e., the emission (absorption) of a  magnon with frequency  $\omega_{qi}$  into (from) the bath. The longitudinal spin susceptibility~\eqref{sz}, instead, encodes  two-magnon (Raman) processes, i.e., a magnon scattering with energy gain (or loss) equal to the frequency $\omega_{qi}$ of the solid-state spin defect~\cite{flebus2018quantum}.
\\

In the long-wavelength limit, the dynamical transverse response of  collinear ferromagnetic systems is  well described by  the Landau–Lifshitz–Gilbert (LLG) equation~\cite{landau1992theory}: 
\begin{eqnarray}
\frac{d\mathbf{s}}{dt}=-\gamma\mathbf{s}\times\mathbf{B}_{eff}+\frac{\alpha}{s}\mathbf{s}\times\frac{d\mathbf{s}}{dt}-\gamma\mathbf{s}\times\mathbf{h}\label{LLGeq},
\end{eqnarray}
where $\alpha$ is the dimensionless Gilbert damping parameter and $\mathbf{B}_{eff}=-\gamma^{-1}\partial \mathcal{H}[\mathbf{s}]/\partial\mathbf{s}$ is the effective magnetic field containing the contributions from an external magnetic field $\mathbf{B}_{0}=B_{0}\hat{\mathbf{z}}$ as well as internal magnetic
fields originating from the interaction of the local spin density with its 
surroundings.  Here we consider a $2d$ ferromagnetic reservoir with  Heisenberg exchange interaction $J$ and uniaxial anisotropy $A$, described by the following Hamiltonian:  
\begin{eqnarray}
\mathcal{H}[\mathbf{s}]&=& \int d^2\mathbf{r}\Big[-J\mathbf{s}(\mathbf{r})\cdot\nabla^2\mathbf{s}(\mathbf{r})-A(\mathbf{s}(\mathbf{r})\cdot\hat{\mathbf{z}})^2\nonumber\\
&&\qquad\quad-\gamma B_0\mathbf{s}(\mathbf{r})\cdot\hat{\mathbf{z}}\Big]\,.\label{bathspec}
\end{eqnarray}
The energy spectrum of the spin waves obeying Eq.~\eqref{bathspec} reads as
\begin{eqnarray}
\omega_F(k) = Dk^2 +\Delta_F,
\label{dispersion}
\end{eqnarray}
where $D =  s J$ represents the spin stiffness and $\Delta_F = 2sA  + \gamma B_0$ is the spin-wave gap.
In the following, we focus on the regime in which one-magnon processes can take place by setting the qubit 
frequency to be larger than the spin-wave gap, i.e., $
\omega_{qi} >\Delta_F$.

 We assume that the magnetic field 
is large enough to align the spin density along its 
direction, i.e., $\textbf{s} =  (s^x, s^y, s)$, with $|
\textbf{s}|\simeq s$, and proceed to linearize Eq.~\eqref{LLGeq} around the equilibrium direction of the magnetic order parameter. Upon Fourier transform, we can recast Eq.~\eqref{LLGeq} in the form of Eq.~\eqref{sminus}, which allows us to identify 
\begin{eqnarray}
\chi^{-+}(\omega, k) = \frac{s/2}{D(k^2 -1/\lambda^2) - i\alpha\omega},
\label{suscepone}
\end{eqnarray}
where  $\lambda$ is a characteristic magnon wavelength defined as $\lambda = \sqrt{D/(\omega - \Delta_F)}$. 
Invoking the relation   $\chi^{+-}(\omega,k)=\left[\chi^{-+}(-\omega,-k)\right]^*$, we find

\begin{eqnarray}
\chi^{+-}(\omega,k)=\frac{ s/2}{D(k^2+1/\lambda'^2)- i\alpha\omega},\label{susceptwo}
\end{eqnarray} 
with $\lambda'=\sqrt{D/(\omega+\Delta_F)}$. The longitudinal spin susceptibility of the magnetic bath can not be captured by the LLG formalism~(\ref{LLGeq}), but depends instead on the pertinent spin transport regime~\cite{kadanoff1963hydrodynamic,flebus2018quantum,fang2022generalized}. 
At wavelengths larger than the magnon mean free path, the dynamics of the longitudinal spin density $s^{z}$ can be treated as diffusive~\cite{wang2022noninvasive,cornelissen2016magnon,flebus2018quantum}, i.e.,
\begin{eqnarray}
\frac{ds^z}{dt}+\nabla\cdot\mathbf{j}_s=-\frac{s^z-\chi_0\gamma h^z}{\tau_s}.\label{zdirecfu}
\end{eqnarray}
Here,  we have introduced the spin-relaxation time $\tau_{s}$ and the spin current $\mathbf{j}_{s}=-\sigma \boldsymbol{\nabla} \mu\,$, where  $\sigma$ is the magnon spin conductivity,  $\mu=   \chi_0^{-1} s^{z} -\gamma h^{z}
$  the chemical potential,  and $\chi_0$ is a scalar quantity corresponding to the static uniform longitudinal susceptibility. 
Recasting Eq.~(\ref{zdirecfu}) in terms of Eq.~(\ref{sz}), one can identify 
\begin{eqnarray}
\chi^{zz} (\omega, k)
=\frac{\chi_0(1+  l^{2}_sk^2)}{-i\omega\tau_s  +(1+ l^{2}_sk^2)},
\label{357}
\end{eqnarray}
where $l_s=\sqrt{\sigma \tau_s/\chi_0}$ is the spin diffusion length. In the regime of interest, i.e., $
\omega_{qi} >\Delta_F$, the process dominating the relaxation dynamics of the  spin defects is the emission (absorption) of a real magnon with frequency $
\omega_{qi} $
 in (from) the magnetic bath~\cite{flebus2018quantum,du2017control,wang2022noninvasive}. Thus, in what follows, we focus on the contributions to the dissipative interactions~\eqref{disscoupling} stemming from the transverse susceptibilities~\eqref{suscepone} and~\eqref{susceptwo}, and neglect the two-magnon processes described by Eq.~(\ref{357}). It is worth noting, however, that several magnetic systems, e.g., most antiferromagnets, display a spin-wave gap $\Delta_F$  far larger than the typical solid-state spin-defect resonance frequency $
\omega_{qi}$. Whether two-magnon scattering processes can mediate quantum cooperative behavior in these systems is an intriguing question  we will address in future investigations.

\section{Master equation }
\label{sec:ME}

We now specialize the master equation~\eqref{densitymatrix} to a qubit array interacting via the ferromagnetic reservoir described in Sec.~\ref{sec:bath}. We focus on the regime in which the energy exchange between the ferromagnetic film and the qubit array is dominated by single-magnon processes, i.e., $\omega_{qi} > \Delta_F$, and neglect the two-magnon (Raman) scattering processes described by Eq.~\eqref{357}. For concreteness, we consider a Yttrium Iron Garnet (YIG) film of thickness $L=20$ nm, spin stiffness $D=5.1\times10^{-28}\;\text{erg}\cdot\text{cm}^2$~\cite{flebus2017magnon}, Gilbert damping parameter $\alpha \sim 10^{-4}$, saturation surface spin density $s=1.2\times10^{-10}\,\ \text{G}^2\,\cdot$\,cm$\,\cdot\,$s and spin-wave gap $\Delta_F=K+ \gamma \text{B}_0$, with zero-field gap $K=3.6\times10^{-18}\;\text{erg}$. As solid-state spin-defects, we focus on an ensemble of NV centers. While an NV center forms a spin-triplet system in the ground state, the degeneracy between $m_s=\pm1$ can be lifted by  applying an external magnetic field $B_0$ along its principal axis.  Each qubit can be then treated as a two-level system with the transition frequency $\omega_{qi}=\Delta_{0}-\tilde{\gamma} B_{0}$, where $\Delta_0=2.87\ \text{GHz}=1.9\times10^{-17}\ \text{erg}$ is the NV-center zero-field splitting. This approximation is valid as long as the thermal energy of the system is significantly smaller than the energy of the higher NV-center transition, i.e., $k_{B}T \ll \Delta_{0}+\tilde{\gamma} B_{0}$, which is consistent with the quantum regime we are focusing on.

We set the distance between the YIG film and the NV-center ensemble to be $d=30$ nm: dissipative interactions between the spin-wave fluctuations of YIG and NV electron spins have already been successfully leveraged in  quantum sensing experiments performed at even larger distances~\cite{du2017control,wang2022noninvasive}. 
The full master equation can be derived by plugging Eqs.~\eqref{suscepone} and~\eqref{susceptwo} into Eqs.~\eqref{dissirealspace},~\eqref{coherealspace} and ~\eqref{dephaserealspace}. In the limit of weak Gilbert damping, i.e., $\alpha \ll 1$, the master equation~\eqref{densitymatrix} reduces to 
\begin{eqnarray}
\frac{d\rho}{dt}&=& -i \left[\mathcal{H}_s+ \sum_{
\alpha\neq\beta}\left(J_{\alpha\beta}\sigma_\alpha^+\sigma_\beta^- + J^{z}_{\alpha\beta}\sigma_\alpha^z \sigma_\beta^z\right),\rho \right]\nonumber\\
&&+\sum_{\alpha\beta}\Gamma_{\alpha\beta}\left(\sigma_\beta^+\rho\sigma_\alpha^--\frac{1}{2}\{\sigma_\alpha^-\sigma_\beta^+,\rho\}\right)\nonumber\\
&&+\sum_{\alpha\beta}\tilde{\Gamma}_{\alpha\beta}\left(\sigma_\beta^-\rho\sigma_\alpha^+-\frac{1}{2}\{\sigma_\alpha^+\sigma_\beta^-,\rho\}\right)\,, \label{masterfull1}
\end{eqnarray}
with
\begin{align}
J_{\alpha\beta}/\nu=&\underbrace{-\frac{\omega_{qi}-\Delta_F}{\Delta_0} P.V.\int_0^\infty d\xi  \; \xi^3 \frac{J_0(\xi\frac{\rho_{\alpha\beta}}{\lambda})e^{-\frac{2d}{\lambda}\xi}}{\xi^2-1}}_{J_{\alpha\beta;1}/\nu} \nonumber \\
& \underbrace{-\frac{\omega_{qi}+\Delta_F}{\Delta_0}\int_0^\infty d\xi \; \xi^3\frac{J_0(\xi\frac{\rho_{\alpha\beta}}{\lambda'})e^{-\frac{2d}{\lambda'}\xi}}{\xi^2+1}}_{J_{\alpha\beta;2}/\nu},\label{coherentJnondissvs}
\end{align}
\begin{align}
J_{\alpha\beta}^z/\nu
=&-\frac{\Delta_F}{\Delta_0}\int_0^\infty d\xi \; \xi^3 \frac{e^{-\frac{2d}{\lambda_{exc}}\xi}J_0(\frac{\rho_{\alpha\beta}}{\lambda_{exc}}\xi)}{\xi^2+1},\label{coherentJznondissvs}\\
\Gamma_{\alpha\beta}/\nu
=& \pi \left[ n_B(\omega_{qi})+1\right]\frac{\omega_{qi}-\Delta_F}{\Delta_0}
J_0\left(\frac{\rho_{\alpha\beta}}{\lambda}\right) e^{-\frac{2d}{\lambda}}\,,\label{dissipativeAnondissvs}
\end{align}
where $P.V.$ stands for principal value.
Here we have introduced $\lambda_{exc}=\sqrt{D/\Delta_F}$, and  $\nu=\frac{\pi h^3(\gamma\tilde{\gamma})^2s\Delta_0}{D^2}$ is a characteristic frequency controlling the strength of coherent and dissipative interactions. The dissipative channels in Eq.~\eqref{masterfull1} contain both correlated emission and correlated pump channels: the probability that a NV center emits (absorbs) a magnon influences the occurrence of an analog event for another NV center at a distance set by the spatial structure of the $\Gamma$ ($\tilde{\Gamma}$) factors in~\eqref{masterfull1}.   This is at variance with quantum optics platforms where only the former can naturally occur~\cite{reitz2022cooperative} or be engineered with Raman sidebands~\cite{seetharam2022correlation}. The  occurrence of a novel form of correlated dissipation is one of the extra leverages offered by solid-state platforms, showing that the cooperative mechanisms expounded in this study transcend a mere one-to-one correspondence with quantum optics.

\begin{figure}[t!]
\includegraphics[width=0.8\linewidth]{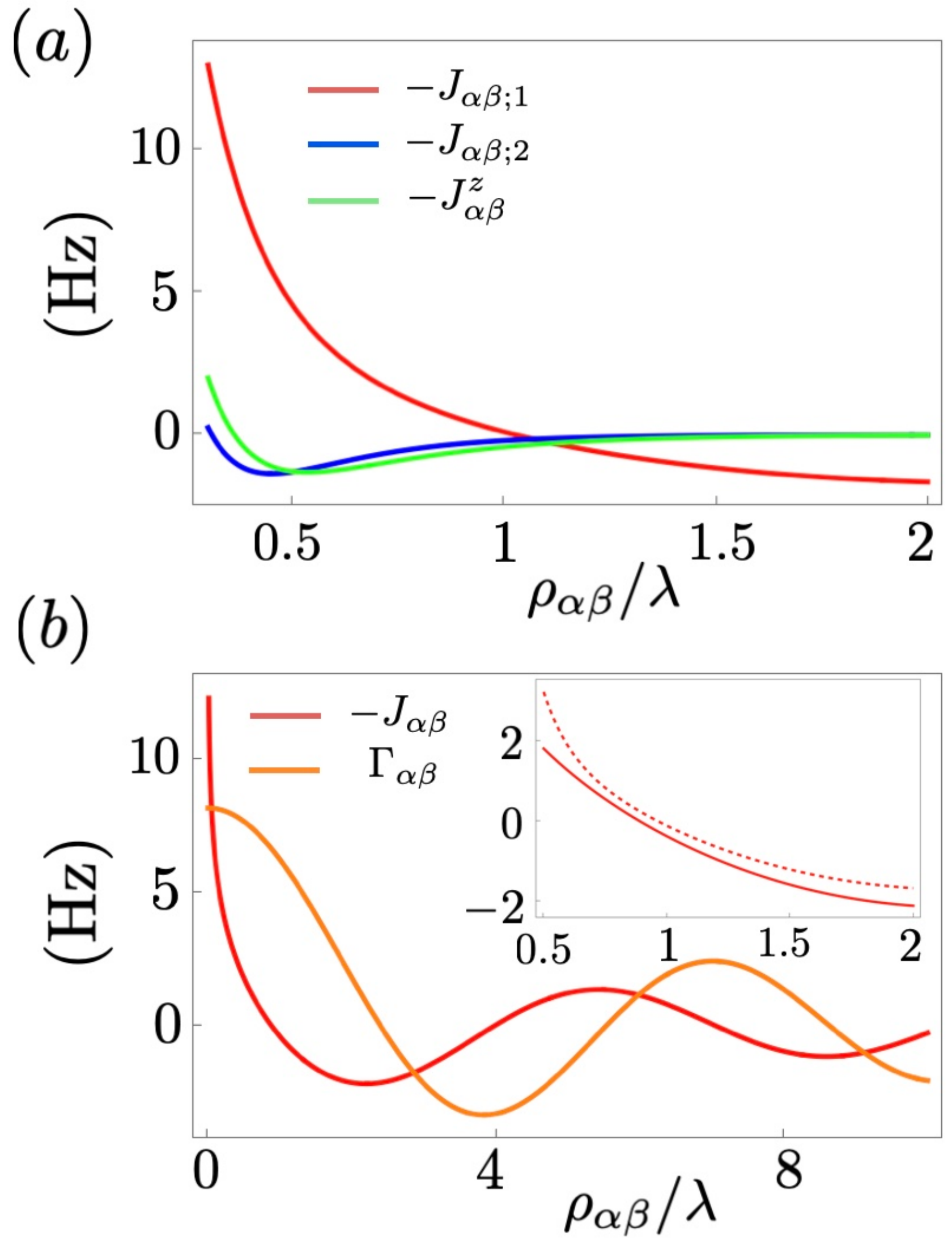}
    \caption{(a) The green, red and blue curves display the dependence of, respectively, the coherent Ising coupling $-J^z_{\alpha\beta}$~ \eqref{coherentJznondissvs},  the first and second  terms on the right-hand side of Eq.  \eqref{coherentJnondissvs}, $-J_{\alpha\beta;1}$ and $-J_{\alpha\beta;2}$, on the interqubit separation $\rho_{\alpha \beta}$  (in units of the wavelength $\lambda$). (b) The red and orange curves show the dependence of,  respectively, the  coherent $-J_{\alpha \beta}$~\eqref{newJ} and incoherent $\Gamma_{\alpha \beta}$~\eqref{dissipativeAnondissvs} couplings on the interqubit separation $\rho_{\alpha \beta}$. Inset: comparison between  Eq.~\eqref{coherentJnondissvs} (red dashed curve) and the approximation provided by Eq.~\eqref{newJ} (red solid curve). In each figure, the parameters used are for an ensemble of NV-center spins interacting via a YIG thin film at  $T=0$.}
    \label{Fig2}
\end{figure}

An ultimate goal would be to realize a quantum hybrid platform in which cooperative quantum dynamics survives at interqubit distances that allow for individual qubit addressability.
Equation~\eqref{dissipativeAnondissvs} shows that the strength of dissipative interactions is controlled by the characteristic wavelength $\lambda = \sqrt{D/(\omega_{qi} - \Delta_F)}$, which can be maximized by choosing a material with large spin stiffness $D$, i.e., high Curie temperature $T_{c}$. However, it is important to note that the strength of interactions scales $\nu \propto D^{-2}$, i.e., the spin-wave fluctuations to which qubits can couple become weaker in stiffer materials. Alternatively, one can minimize the detuning between the NV-center resonance frequency $\omega_{qi}$ and the spin-wave gap $\Delta_{F}$ via an external magnetic field $B_{0}$, i.e., $\omega_{qi}-\Delta_{F}=\Delta_{0}- K-2\gamma B_{0}$. By choosing $B_0\sim 40$ mT, we set  $\omega_{qi}-\Delta_{F} \sim 100$ MHz, consistently with the experimental frequency resolution limits. In this regime, we find $\lambda 
\sim 280$ nm and $\lambda', \lambda_{exc} 
\sim 50$ nm.

Our proposed setup fully satisfies the Markovian approximation used in the derivation of our framework, which is also invoked by Ref.~\cite{fukami2021opportunities} in the study of the dynamics of two NV centers interacting via a YIG waveguide. The lifetime of the relevant magnonic excitations can be estimated from the linearized LLG equation~(\ref{LLGeq}) at $k_0=\lambda^{-1}$ as $\tau_{0} = (2 \alpha \omega_{k_0})^{-1} \sim 1 \; \mu \text{s}$ for high-quality YIG samples with $\alpha \sim 10^{-4}$. For $T\rightarrow 0$ and $e^{-\frac{2d}{\lambda}}\rightarrow 1$,   the spontaneous decay rate of an isolated NV center interacting with the magnetic bath can be recasted from Eq.~(\ref{dissipativeAnondissvs}) as $\Gamma_0=\frac{\pi^2h^3 (\gamma \tilde{\gamma})^2s(\omega_{qi}-\Delta_F)}{D^2} \sim 10$ Hz.  Although this is larger than the intrinsic sub-Kelvin relaxation rates~\cite{cambria2023temperature}, it still corresponds to a lifetime far exceeding that of the bath excitations by several orders of magnitude. Even at higher temperatures, e.g., $T=100$ mK, we find that significant timescale separation between the qubit and bath dynamics that is at the core of the Markov approximation still holds, i.e., $\Gamma_{0} \sim 10$ Hz. It is also essential to note that here we focus on coupling NV centers to a continuum of propagating magnons, which, in principle, enables Markovian dynamics even when the magnons and the NV centers  lifetimes are comparable. The only significant deviation from this behavior would occur if the system's boundaries cause the magnons to reflect with high probability, directing them back toward the NV centers before dissipating. Having this knowledge, system sizes and boundary conditions can easily be engineered to avoid this effect. We also find that  retardation effects in the effective inter-qubit couplings can be also safely neglected as the time taken for a real magnon with the velocity $v_{magn}$, to travel a wavelength $\lambda$, approximately $t_1 \simeq  \lambda/v_{magn} \sim 1$ ns, is significantly shorter than the NV center relaxation time $\Gamma^{-1}_0$.   The fact that $t_1/\Gamma^{-1}_0$ serves as an expansion parameter for the effects of retardation is established, for example, in Ref. \cite{guimond2016rabi}.

\begin{figure*}[htbp]
    \centering
\includegraphics[width=1\linewidth]{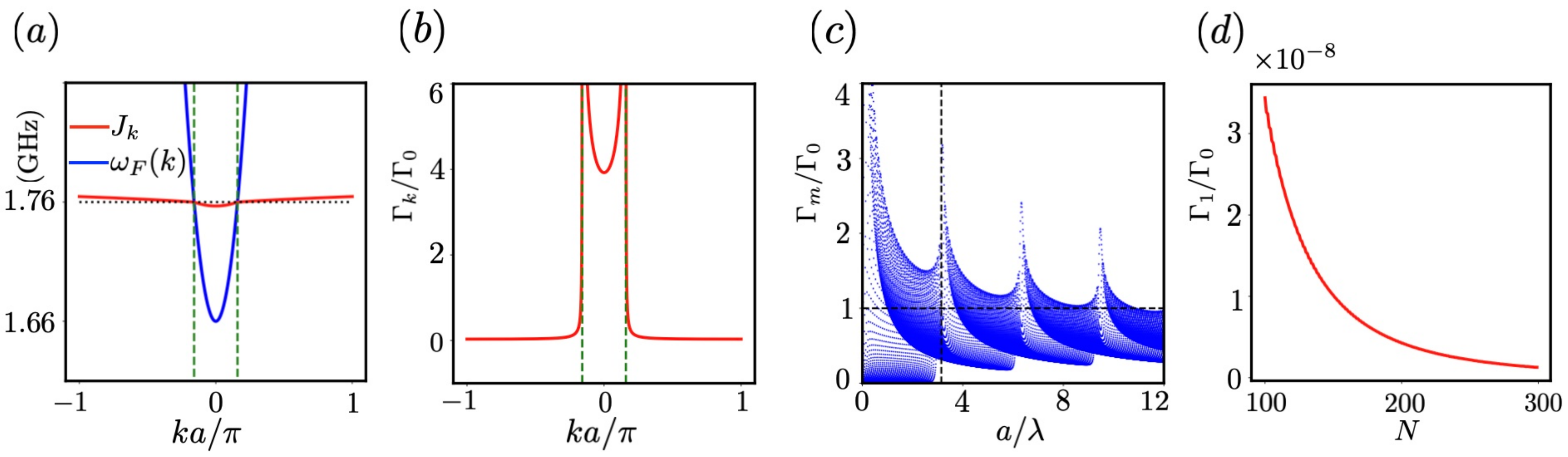}
    \caption{(a) Dispersion relation $J_{k}+\omega_{qi}$ for the single-excitation modes of an infinite, one-dimensional chain (red line). The blue line represents the spin-wave dispersion $\omega_{F}(k)$ of the magnetic film~\eqref{dispersion}, while the black dashed line corresponds to the resonance frequency $\omega_{qi}$ of the qubits. (b) Collective decay rate $\Gamma_{k}$ of the qubit array normalized by the relaxation rate $\Gamma_{0}$ of a noninteracting array. The green dashed lines separate subradiant from superradiant modes. (c) Normalized decay rates $\Gamma_{m}$ at different lattice constants $a/\lambda$ for a finite array of $N=40$ solid-state spin defects. Subradiant modes with relaxation rates approaching zero arise only for $a/\lambda < \pi $ (vertical dashed line), in agreement with the infinite chain analysis. (d) The decay rate of the most subradiant eigenmode as a function of the number of qubits for $a/\lambda=0.1$.  In each figure, the parameters used are for an ensemble of NV-center spins interacting via a YIG thin film.}
    \label{Fig22}
\end{figure*}

Figure~\ref{Fig2}(a) shows clearly that spatial decay of the  coherent Ising coupling~\eqref{coherentJznondissvs} (green curve), which is controlled by the wavelength $\lambda_{exc}$, is very rapid compared to the first term  on the right-side of Eq.~\eqref{coherentJnondissvs} (red curve). Therefore, we set $J^{z}=0$ as it does not play a substantial role in the quantum dynamics we are interested in.
Figure~\ref{Fig2}(a)  shows also that the strength of the first term  on the right-side of Eq.~\eqref{coherentJnondissvs} (red curve), whose decay is controlled by the wavelength $\lambda$, is much larger than the second term on the right-hand side of Eq.~\eqref{coherentJnondissvs} (blue curve), which scales with the wavelength $\lambda'$. We therefore retain only the first term  on the right-side of Eq.~\eqref{coherentJnondissvs} and  take the limit $e^{-\frac{2d}{\lambda}}
 \rightarrow 1$, which allows to rewrite Eq.~\eqref{coherentJnondissvs} as
 \begin{align}
J_{\alpha\beta}/\nu&=\frac{\pi}{2} \frac{\omega_{qi}-\Delta_{F}}{\Delta_0} Y_0\left(\frac{\rho_{\alpha\beta}}{\lambda}\right)\,.
\label{newJ}
 \end{align}
As shown by the inset of Fig.~\ref{Fig2}(b), Eq.~\eqref{newJ} (solid red curve) is a good approximation of Eq.~\eqref{coherentJnondissvs} (dashed red curve)  within the selected parameter regime.
Therefore, Eq.~\eqref{masterfull1} can be rewritten as
 \begin{eqnarray}
\frac{d\rho}{dt}&=& -i \left[ \mathcal{H}_s+\sum_{
\alpha\neq\beta}J_{\alpha\beta}\sigma_\alpha^+\sigma_\beta^-,\;\rho \right]\nonumber\\
&&+\sum_{\alpha\beta}\Gamma_{\alpha\beta}\left(\sigma_\beta^+\rho\sigma_\alpha^--\frac{1}{2}\{\sigma_\alpha^-\sigma_\beta^+,\rho\}\right)\nonumber\\
&&+\sum_{\alpha\beta}\tilde{\Gamma}_{\alpha\beta}\left(\sigma_\beta^-\rho\sigma_\alpha^+-\frac{1}{2}\{\sigma_\alpha^+\sigma_\beta^-,\rho\}\right),\label{masterfull}
\end{eqnarray}
where the strength of dissipative and coherent interactions is given, respectively, by Eqs.~\eqref{dissipativeAnondissvs} and~\eqref{newJ}, while thermal equilibrium requires $\tilde{\Gamma}_{\alpha\beta}=e^{-\beta h\omega_{\text{qi}}}\Gamma_{\alpha\beta}$.

Figure~\ref{Fig2}(b) shows that, in the regime $\omega_{qi} > \Delta_{F}$, the strength of the qubit-qubit interaction $\Gamma_{\alpha \beta}$ mediated by dissipative qubit-bath interactions, i.e., exchange of real magnons, is comparable to the coherent exchange interaction  $J_{\alpha\beta}$ mediated by virtual magnon processes.

\section{Single-excitation sector}
\label{sec:single}

In this section, following a standard procedure used to study correlated emission in atomic ensembles~\cite{albrecht2019subradiant,asenjo2017exponential}, we develop analytical insights into our model by analyzing its single-excitation sector in the zero-temperature limit. The impact of finite temperature will be discussed in Sec.~\ref{sec:thermal}.
   For vanishing temperature, i.e., $T\rightarrow 0$, the master equation~\eqref{masterfull} can be written as 
\begin{eqnarray}
\frac{d \rho}{dt}=-i\left( \mathcal{H}_{eff}\rho-\rho\mathcal{H}_{eff}^\dag \right)+\sum_{\alpha\beta}\Gamma_{\alpha\beta}\sigma_\beta^+\rho\sigma_\alpha^-,\label{newmaster}
\end{eqnarray}
where we have introduced the effective non-Hermitian Hamiltonian defined as $\mathcal{H}_{eff}=\mathcal{H}_s+\mathcal{H}+\mathcal{H}_{nh}$ with $\mathcal{H}_{nh}=-i\sum_{\alpha\beta}(\Gamma_{\alpha\beta}/2)\sigma_\alpha^-\sigma_\beta^+$. The single-excitation sector corresponds to a subspace of the full Hilbert space where there is only up to a single excitation shared in the whole array. The dynamics in this state are well-described by an effective Hamiltonian, allowing us to neglect the quantum jump terms, i.e., the second term on the right-hand side of Eq.~(\ref{newmaster})~\cite{gardiner2004quantum}.  This suggests that the proposed quantum hybrid spin systems could also serve as a novel platform for exploring non-Hermitian magnonic phenomena~\cite{hurst2022non,li2023reciprocal,gunnink2022nonlinear,kamboj2023oscillatory,deng2023exceptional,lee2015macroscopic,liu2019observation,li2022multitude} as well as their resilience against many-body quantum effects~\cite{minganti2019quantum}.

In our analysis, we first consider an infinite chain of solid-state spin defects:  even though this scenario is not realistic, it will allow us to gain key insights into the physical properties of the system. The single excitation is not localized at a single site, but is coherently distributed across all sites, thereby forming a spin wave with creation operator $S^\dag_k=(1/\sqrt{N})\sum_{\alpha}e^{-ikr_\alpha}\sigma_\alpha^-$, where $k$ denotes the quasi-momentum.
In this representation, the effective Hamiltonian takes the form:
\begin{eqnarray}
\mathcal{H}_{eff}(k) = \sum_{k}\left(J_k-i\Gamma_k/2\right) S^\dag_{k}S_{k},\label{effhamqspace}
\end{eqnarray}
where the coefficients $J_k$ and $\Gamma_k$ are defined as
\begin{eqnarray}
J_k &=& \sum_{n=\pm 1}^{\pm\infty}J(|n|a)e^{-inka}, \label{453}\\
\Gamma_k &=& \sum_{n=0}^{\pm\infty}\Gamma(|n|a)e^{-inka} \label{454},
\end{eqnarray}
where $n$ are  non-negative integers and $a$ the lattice constant of the qubit array.
Here, $J_k$ and $\Gamma_k$ can be identified as, respectively, the frequency shift  with respect to the isolated qubit frequency $\omega_{qi}$ and the single-excitation decay rate of mode  $k$. 

Figures~\ref{Fig22}(a) and (b)  display the dependence of the collective excitation spectrum $J_k$ and the decay rate  $\Gamma_k$ on wavevectors within the first Brillouin zone. We observe that the single-excitation modes of the array near the center of the Brillouin zone, i.e., within the green dashed lines in Fig.~\ref{Fig22}(a), have quasi-momenta along the chain direction that are smaller than the maximum momentum of the spin waves in the magnetic film at the same energy. As a result, these modes can strongly couple to the spin waves of the reservoir, leading to rapid decay characterized by a relaxation rate $\Gamma_{k}$ that is larger than that of an isolated qubit, as shown by Fig.~\ref{Fig22}(b). This behavior is symptomatic of superradiance.
On the other hand, modes with quasi-momenta greater than the momentum of the magnon modes are completely decoupled from the magnons. These modes are unable to radiate away energy through the bath, resulting in long-lived, subradiant states. An akin effect has been noted for atoms interacting with light  -- specifically, spin wavevectors whose magnitudes exceed the dispersion relation of light, i.e., $|k|>\omega/c$ with $c$ being the light speed, have exactly zero decay rate in the limit of an infinite chain~\cite{asenjo2017exponential,albrecht2019subradiant}.

For subradiant states to appear,  the spin-wave frequency $\omega_F(k)$~\eqref{dispersion} in the magnetic film at certain wavevectors should be larger than the qubit resonance frequency $\omega_{qi}$, i.e.,  $\omega_{qi} < \omega_F(k)$. At these wavevectors, the energy emitted from the qubits is not sufficient to create single magnons in the magnetic film.
This condition must be satisfied within  the boundaries of the first Brillouin zone: $\omega_F(\pi/a) = D(\pi/a)^2 + \Delta_F \geq \omega_{qi}$, which translates into the constraint  $a/\lambda \leq \pi$ for the characteristic wavelength $\lambda$. 

%The wavelength $\lambda = \left[D/(\omega_{qi} - \Delta_F)\right]^{1/2}$ can be enhanced by controlling the detuning $\omega_{qi} - \Delta_F$ via an external magnetic field and by choosing a magnetic bath with high Curie temperature. 

In the following, we investigate the properties of the single-excitation sector of a qubit array of finite size. By utilizing the property that the non-Hermitian Hamiltonian $\mathcal{H}_{eff}$ commutes with the particle number operator $\hat{n}=\sum_{\alpha}\sigma_\alpha^+\sigma_\alpha^-$, i.e., $\hat{n} \mathcal{H}_{eff}-\mathcal{H}_{eff}\hat{n}=0$, one can characterize its eigenstates  within each excitation manifold. 
To this end, we expand $\mathcal{H}_{eff}$ in the basis $\{\ket{\phi_\alpha}\}$, such that $\mathcal{H}_{eff,\alpha\beta}=\bra{\phi_\alpha}\mathcal{H}_{eff}\ket{\phi_\beta}$, where $\ket{\phi_\alpha}=\ket{\uparrow,\cdots,\uparrow,\downarrow_\alpha,\uparrow,\cdots,\uparrow}$ represents the state in which only the $\alpha$th spin is excited. Next, by diagonalizing the matrix $\mathcal{H}_{eff}$, we obtain a set of eigenvalues $\{E_m\}$, with 
$1\leq\,m\leq\,N$, whose imaginary part corresponds to the decay rate of the single-excitation modes. 
Figure~\ref{Fig22}(c) shows the decay rates of  the single-excitation modes of a chain of $N=40$  solid-state spin defects as a function of the (renormalized) lattice constant $a/\lambda$. As the interqubit distance increases, the effective qubit-qubit dissipative interaction decreases and the collective decay rates approach the spontaneous emission rate of a single qubit. 
Conversely, for smaller lattice constants, one can observe a reduction in the minimum decay rates, which indicates the formation of subradiant states.
Remarkably, when the interparticle distance reaches the critical point $a/\lambda=\pi$, the most subradiant decay rate experiences a significant drop, approaching zero as suggested by the infinite array analysis.
Figure~\ref{Fig22}(d) shows the decay rate $\Gamma_1$ of the most subradiant state $|\psi_1^R\rangle$ as a function of the number $N$ of solid-state spin defects. For an interqubit distance $a/\lambda=0.1$ the relaxation rate of the most subradiant state approaches quickly zero even for a finite chain.

\section{Multi-excitation sector}
\label{sec:multi}

\begin{figure*}[htbp]
    \centering
\includegraphics[width=1\linewidth]{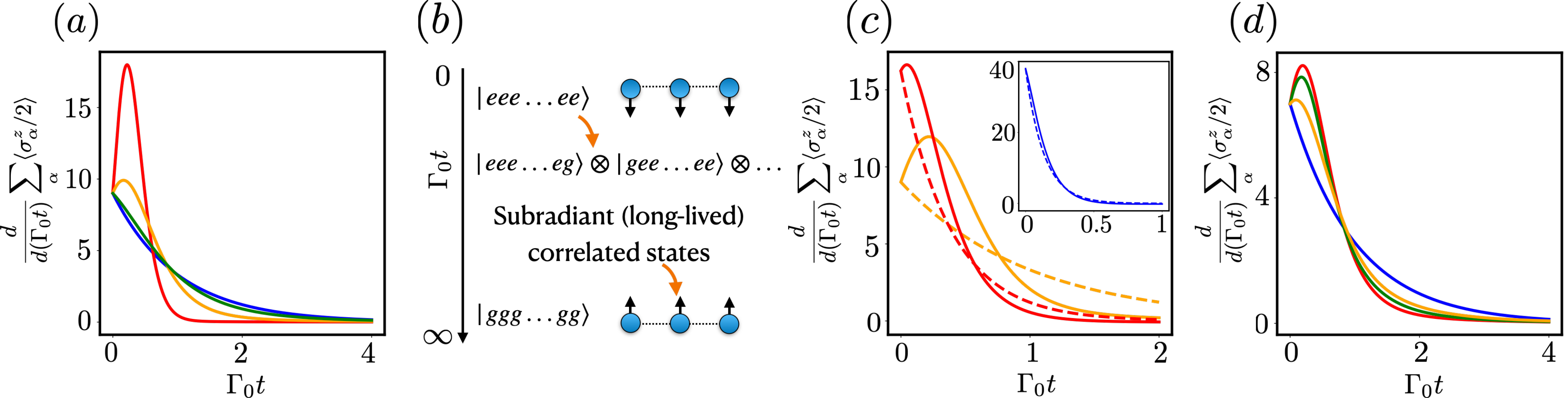}
    \caption{(a) Zero-temperature collective relaxation rate of an ensemble of $N=9$ solid-state spin defects initialized in the excited state as a function of time. The  red, orange and green lines correspond to arrays with lattice constant $a/\lambda=0.1, 0.7$ and $2$, respectively. The blue line corresponds to magnon emission rate of a noninteracting ensemble.  When the spin defects are placed sufficiently close to each other, the collective emission rate displays a superradiant burst followed by a subradiant tail. (b) The ensemble relaxation dynamics goes through the ``Dicke ladder", i.e., the energy levels of the many-body system can be organized into a ladder-like structure, where each rung corresponds to a given number of excitations. Superradiant states decay toward subradiant states, which can exhibit strong quantum correlations and entanglement. (c)  Collective relaxation rate of an ensemble of $N=9$ solid-state spin defects initialized in the excited state as a function of time  at $T=10$ mK (orange curves), $T=100$ mK (red curves) and $T=300$ mK (blue curves in the inset). The solid and dashed lines correspond to the relaxation rates  calculated for  $a/\lambda=0.4$ and $ a/\lambda \rightarrow \infty$ (i.e., a noninteracting array), respectively. (d) Zero-temperature collective relaxation rate of an ensemble of $N=7$ solid-state spin defects initialized in the excited state as a function of time for $a/\lambda=0.5$ and disorder strength $\xi=0$ (red line), $\xi=5$ (green line) and $\xi=10$ (orange line). The blue line corresponds to a noninteracting ensemble.  (a,c,d) The time $t$ is in units of the relaxation time $\Gamma_{0}$ of an array of $N$ noninteracting spin defects.  In each figure, the parameters used are for an ensemble of NV-center spins interacting via a YIG thin film.}
    \label{Fig55}
\end{figure*}
In this section, we explore the ensemble dynamics in the multi-excitation sector, i.e., corresponding to multiple qubits simultaneously initialized into the excited state. This regime is immediately relevant to experimental verifications of our proposed model, as it does not require single-qubit addressability. We show that, at temperatures low enough, the collective relaxation rates routinely measured in quantum sensing setups can display clear signatures of cooperative quantum behavior. Finally, we investigate their robustness against experimentally relevant perturbations, i.e., temperature and spatial disorder.

\subsection{Zero-temperature dynamics}

In the multi-excitation sector, the quantum jump terms can not be neglected, and, thus, Eq.~\eqref{masterfull}  has to be solved exactly. 
We first consider  the zero-temperature limit, i.e.,   $\tilde{\Gamma}_{\alpha \beta}=0$, and solve Eq.~\eqref{masterfull} numerically using the quantum jump method~\cite{johansson2012qutip}. We compute the ensemble collective relaxation rate, i.e., $ \frac{d}{d(\Gamma_0t)}\sum_{\alpha=1}^{N}  \langle \sigma^{z}_{\alpha}/2 \rangle $, which is  proportional to the photoluminescence signal probed in NV-center relaxometry measurements~\cite{casola2018probing}.

Figure \ref{Fig55}(a) displays the time evolution of the collective relaxation rate of an array of $N=9$ solid-state spin defects initialized in the fully excited state. When the inter-qubit distance is smaller than the characteristic wavelength, i.e.,  $a\lesssim\lambda$, the emission rates of the correlated ensemble (red and orange lines) are sharply different from the uncorrelated case (blue line). At short times, the relaxation rate of the correlated ensemble displays a burst, in contrast to the exponential decay of the uncorrelated ensemble. Comparing the case of $a=0.7\lambda$ (orange curve) and $a=0.1\lambda$ (red curve), we find that the intensity of the burst is enhanced for smaller inter-qubit distances, while the time needed to reach the superradiance peak becomes shorter. The dynamics of the many-body density matrix can be understood in terms of the ``Dicke ladder" shown in Fig.~\ref{Fig55}(b).
 The superradiant speed up of the collective relaxation, which removes the excess energy from the ensemble in a semiclassical fashion, is followed by a slower transient dynamics of subradiant modes that might be a source of quantum correlations~\cite{asenjo2017exponential, masson2020many,rui2020subradiant}. 

Our results suggest that signatures of superradiance and subradiance are experimentally detectable  even for a small, i.e., $N=9$, ensemble of correlated solid-state spin defects,  as long as the ensemble relaxation rate is larger  than the  intrinsic relaxation time of solid-state spin defects. Since the latter can reach even minutes at very low temperatures~\cite{jarmola2012temperature}, the 
strength $\nu \propto D^{-2}$ of the interactions between a spin defect and a ferromagnetic bath does not have to be particularly large, which leaves room for maximizing the characteristic lengthscale $\lambda \propto D^{1/2}$ via a 
stiffer magnetic bath.

\subsection{Thermal fluctuations, spatial disorder and inhomogeneous broadening}
\label{sec:thermal}

In our setup, the role of finite-temperature baths opens up a line of investigation that has not been systematically explored in  light-matter interfaces. At finite temperatures, a magnetic bath at thermal equilibrium is populated according to the Bose-Einstein distribution, and thermal magnons  can interact with the qubit ensemble via the correlated absorption term $\tilde{\Gamma}_{\alpha\beta}$ in Eq.~\eqref{masterfull}.

We explore the impact of thermal fluctuations on quantum cooperativity by numerically simulating Eq.~\eqref{masterfull} at finite temperature $T$.  Figure~\ref{Fig55}(c) shows that cooperative   behavior is suppressed with increasing temperature, as one would naturally expect since the pumping term counterbalances the collective decay through the Dicke ladder (cf. Fig.~\ref{Fig55}) which is responsible for superradiance. However, for  temperatures of the order $T \sim \omega_{qi}/k_{B}$, i.e., $T=100$ mK, which can be feasibly accessed experimentally,  superradiant and subradiant deviations from the exponential relaxation of an isolated ensemble (red line) are still clearly visible. Interestingly, when the temperature is finite but not high enough to suppress cooperative effects, the superradiant burst is reached at shorter times. This occurs as the temperature increases the dissipative correlations according to Eq.~\eqref{dissipativeAnondissvs}, which leads to stimulated emission and a consequent  speed-up in the collective relaxation dynamics.

Although the fabrication efficiency of ordered arrays of solid-state spin defects is continuously improving ~\cite{wolfowicz2021quantum,spinicelli2011engineered,tsukanov2013quantum} and solid-state spin defects in hexagonal boron nitride (hBN) can be deterministically generated~\cite{liu2022spin}, spatial disorder in the qubit positions is still likely to play a role in the experimental realizations of our proposal. 
In order to model spatial disorder we consider an array 
with random variations  in the individual position $r_{\alpha}$ of each 
qubit, i.e., $r_{\alpha} \rightarrow r_{\alpha} +  \delta 
r_{\alpha}$. We treat the  positional displacement $\delta 
 r_{\alpha}$ as a random variable uniformly distributed 
 within the interval $\delta r_{\alpha} \in \left( -a \xi /2, a \xi /2 \right)$  where $\xi$ quantifies the strength 
of disorder. We consider $\xi=5,10$ and for each  degree 
of disorder we report an average of the collective relaxation rate over $N=10$ realizations of qubit positions. Figure~\ref{Fig55}(d) shows that signatures of cooperative behavior are still clearly visible for large disorder strength $\xi$, demonstrating the existence of super/subradiant dynamics under a wide range of settings. This aligns with other work demonstrating cooperative dynamics under very different circumstances ~\cite{bradac2017room,inouye1999superradiant,ferioli2021laser,raino2018superfluorescence,scheibner2007superradiance,cardenas2023many}. On the other hand, we expect that if one wants to engineer specific many-body states using cooperative dynamics, the tolerance to disorder will be more stringent.

It is also important to note that our results are constrained to  the relatively small number $N$ of qubits that quantum many-body simulations can numerically tackle.  Since these phenomena are cooperative, the intensity of both subradiant and superradiant signals increases with the number of solid-state defects~\cite{breuer2002theory, masson2020many}, which might further relax the temperature and disorder constraints.

 Due to the constraint on the size of the system set by numerical simulations, we  do not include explicitly in our model the inhomogeneous broadening that unavoidably arises in a NV-center ensemble due to the random distribution of NV symmetry axes, excess nitrogen P1 centers, uncharged NV0, and nuclear spins. However, a comparison between our results and the experimental work of Angerer \textit{et al.}~\cite{angerer2018superradiant} suggest that signature of superradiant and subradiant dynamics should be observable in a dense, disordered ensemble of NV centers interacting with YIG film cooled at low temperatures, even in the presence of significant broadening in the NV-center ensemble. In Ref.~\cite{angerer2018superradiant}, the authors probe the relaxation dynamics of an ensemble of NV centers interacting dissipatively via a three-dimensional lumped element resonator in the quantum regime. While the  inhomogeneous broadening $\Delta \omega$ in the resonance frequencies of the NV-center ensemble is much larger than the bath-qubit coupling quantified by the Purcell factor $\Gamma_0$ (i.e., $\Delta \omega \sim \text{MHz}$ and $\Gamma_0 \sim 10^{-9} \; \text{Hz}$),  the qubit ensemble emits a superradiant pulse a trillion of times faster than the decay for an individual NV centre. This enhancement arises from the large number $N$ of emitters: NV centers packed within a region of radius $\lambda$ act as a single emission channel ($L=S^-$) with total spin $N/2$. Consequently, they effectively enhance the dissipation rate by a factor that is extensive in system size, granting robustness against inhomogeneous broadening. This phenomenon exemplifies a fundamental \textit{many-body} advantage of our proposal, setting it apart from previous works focusing solely on two-qubit gate architectures, which require identical qubits to achieve large cooperativities.
For our proposed setup, we find that the Purcell factor exceeds that of Ref.~\cite{angerer2018superradiant} by several orders of magnitude, i.e., $\Gamma_{0} \sim 10$   Hz for $T \sim 0-100$ mK, suggesting that engineering robust correlated emission might not require particularly dense ensembles.

\section{Conclusions and perspectives}
\label{sec:concl}

In this work, we develop a comprehensive formalism for exploring the quantum many-body dynamics of an array of solid-state spin defects interacting dissipatively with the electromagnetic noise emitted by a common solid-state reservoir. 
To assess the experimental significance of our proposal, we specialize our findings to an NV-center array coupled to a ferromagnetic bath and show that signatures of cooperative quantum dynamics, i.e., superradiant and subradiant states, can be engineered within experimentally accessible regimes. We also demonstrate the robustness of these phenomena to perturbations  present in any practical implementation. 

Our framework captures the dependence of the qubit dynamics on the cross-correlation spectra of the  noise generated by the solid-state bath, thereby laying the foundation for a multi-qubit approach to sensing of quantum materials. It is natural to anticipate that this novel approach might allow probe  properties invisible to the current state-of-the-art spectroscopic techniques based on single-qubit sensors, such as the scaling of correlation lengths and entanglement that play a central role in several  condensed matter phenomena.

Our results also open up a novel, dissipation-driven route to engineer robust many-body entanglement among solid-state qubits, which can find applications ranging from one-way quantum computer  schemes~\cite{nielsen2006cluster,briegel2009measurement} to high-precision quantum metrology~\cite{wineland1992spin,ostermann2013protected}.  Due to the diverse and tunable properties of solid-state systems, our work unlocks an unprecedented versatility in reservoir engineering that future investigations should systematically address to identify the optimal bath and experimental regimes for generating robust, long-range correlations between solid-state spin defects.

The tunability and variety of solid-state baths pave also the way for the exploration of quantum cooperative phenomena in platforms with functionalities distinct from quantum optics setups.  The dynamics of magnons and topological spin textures, for example, can be  guided through the influence of diverse non-equilibrium agents, such as currents and gradients in chemical potential, thus unlocking the potential for cooperative phenomena in the presence of  controllable far-from-equilibrium baths. This capability remains beyond the grasp of quantum optics, where the environment assumes a passive role as an inert mediator of interactions.
It also constitutes a rich avenue of exploration from a methodological standpoint. Solving  a 1$d$ (or more realistically 2$d$) quantum magnet with correlated dissipation mediated by an active quantum bath is beyond the reach of Lindblad equation, due to the potential breakdown of Markovianity. Furthermore, several heavy-numerics based methods would fail to track the long-time dynamics, even for a few qubits ensemble, due to memory overheads. The few remaining options could consist  in relying for short-time dynamics on semi-classical phase space methods~\cite{schachenmayer2015many} or developing from scratch non-equilibrium diagrammatics for non-unitary, non-Markovian systems~\cite{stefanucci2024kadanoff} (see Refs.~\cite{schuckert2018nonequilibrium,babadi2015far} for the unitary case, or Refs.~\cite{hosseinabadi2023nonequilibrium,hosseinabadi2023dynamics} for cavity QED distant relatives of the situation described here). This appears as a challenging,  yet extremely rewarding, line of research which we aim to follow up in the near future.

The quantum many-body dynamics emerging in our solid-state setup  exhibits a unique interplay of correlated absorption, emission, and dephasing, setting it apart from conventional quantum optics where correlated emission, including super- and subradiance, is the main manifestation of structured dissipation. While proposals exist to achieve correlated emission with arbitrary spatial profiles~\cite{hung2016quantum,seetharam2022dynamical,seetharam2022correlation,marino2022universality}, engineering other forms of non-local dissipation remains a   challenge in the field of AMO. In contrast, the solid-state platforms at the center of our work are extremely flexible in this regard, while still benefiting from the advantages offered by conventional dissipative state preparation, such as insensitivity to initial conditions and robustness against imperfections. For instance, our 
experimental setup naturally lends itself to investigating the 
prerequisites for establishing many-particle entanglement in the 
presence of spatially non-local 
dephasing~\cite{seif2022distinguishing} — a captivating and 
counterintuitive avenue hitherto explored from a purely theoretical standpoint.

In summary, our work lays the foundation for merging spintronics and quantum optics towards a common research horizon in the incoming future. We anticipate that this emerging interdisciplinary field will enable the investigation of cooperative quantum phenomena in regimes hardly pursuable in quantum optics setups, while paving the way for novel quantum sensing modalities and reservoir-engineering approaches for generating many-body entangled states in tunable solid-state platforms.

\section{ACKNOWLEDGEMENTS}
The authors thank  Y. Tserkovnyak, K. Holczer, and  C. Du  for helpful discussions and
O. Chelpanova for support with the numerical simulations.  B. Flebus acknowledges support from DOE under Grant. No DE-SC0024090. D.E. Chang acknowledges support from the European Union, under European Research Council grant agreement No 101002107 (NEWSPIN); the Government of Spain under the Severo Ochoa Grant CEX2019-000910-S [MCIN/AEI/10.13039/501100011033]); QuantERA II project QuSiED, co-funded by the European Union Horizon 2020 research and innovation programme (No 101017733) and the Government of Spain (European Union NextGenerationEU/PRTR PCI2022-132945 funded by MCIN/AEI/10.13039/501100011033); Generalitat de Catalunya (CERCA program and AGAUR Project No. 2021 SGR 01442); Fundaci\'{o} Cellex, and Fundaci\'{o} Mir-Puig.  J. Marino acknowledges support by the Deutsche Forschungsgemeinschaft (DFG, German Research Foundation) – Project-ID 429529648 – TRR 306 QuCoLiMa (“Quantum Cooperativity of Light and Matter”),  by the Dynamics and Topology Centre funded by the State of Rhineland Palatinate, and by  the QuantERA II Programme that has received funding from the European Union’s Horizon 2020 research and innovation programme  under Grant Agreement No 101017733 ('QuSiED') and by the DFG (project number 499037529).

\bibliography{library}

\end{document}